\documentclass[%
 reprint,
 amsmath,amssymb,
 aps,
 prx,
]{revtex4-2}

\usepackage{graphicx}
\usepackage{dcolumn}
\usepackage{bm}
\usepackage{subfigure}
\usepackage{threeparttable}
\usepackage{hyperref}

\usepackage{supertabular,booktabs}

\usepackage{amsmath}
\usepackage{amssymb}
\usepackage{mathtools}
\usepackage{amsthm}
\usepackage{listingsutf8}
\usepackage{xcolor}
\usepackage{tcolorbox}
\theoremstyle{plain}
\newtheorem{theorem}{Theorem}

\usepackage{tabularx}
\usepackage{array}
\usepackage{multirow}
\usepackage{seqsplit}
\usepackage{makecell}
\usepackage{hhline}
\usepackage{ltablex}
\usepackage{booktabs}
\usepackage{ragged2e}
\usepackage{makecell}
\usepackage{siunitx}

\usepackage{quantikz}
\usepackage{adjustbox}

\begin{document}
\title{Out-of-Time-Order Correlator Spectroscopy}

\author{Keisuke Fujii}

\affiliation{Graduate School of Engineering Science, The University of Osaka, 1-3 Machikaneyama, Toyonaka, Osaka 560-8531, Japan}
\affiliation{Center for Quantum Information and Quantum Biology,
The University of Osaka, 1-2 Machikaneyama, Osaka 560-0043, Japan}
\affiliation{RIKEN Center for Quantum Computing (RQC), Hirosawa 2-1, Wako, Saitama 351-0198, Japan}

\date{\today}

\begin{abstract}
Out-of-time-order correlators (OTOCs) are central probes of quantum 
scrambling, and their generalizations have recently become key primitives for both benchmarking 
quantum advantage and learning the structure of Hamiltonians.
Yet their behavior has lacked a unified algorithmic interpretation.
We show that higher-order OTOCs naturally fit within the framework of 
quantum signal processing (QSP): each $\mathrm{OTOC}^{(k)}$ measures the 
$2k$-th Fourier component of the phase distribution associated with the 
singular values of a spatially resolved truncated propagator.  
This explains the contrasting sensitivities of time-ordered correlators (TOCs) and higher-order 
OTOCs to causal-cone structure and to chaotic, integrable, or localized 
dynamics.
Based on this understanding, we further generalize higher-order OTOCs 
by polynomial transformation of the singular values of the spatially resolved truncated propagator.
The resultant signal allows us to construct frequency-selective 
filters, which we call \emph{OTOC spectroscopy}. This extends conventional OTOCs into a mode-resolved tool 
for probing scrambling and spectral structure of quantum many-body dynamics.
\end{abstract}

\maketitle

\section{Introduction}
Quantum computers promise computational capabilities that surpass those 
of classical devices, enabling applications ranging from factorization~\cite{shor1994algorithms}, quantum 
simulation~\cite{lloyd1996universal,abrams1999quantum,aspuru2005simulated} and linear system solvers~\cite{harrow2009quantum}.  
As hardware platforms continue to advance, a central challenge has emerged 
in determining how we can reliably characterize, benchmark, and ultimately understand the computational advantage of such quantum processors.

Early demonstrations, such as quantum computational supremacy experiments,
have provided compelling evidence that quantum devices 
can generate output distributions intractable for classical 
supercomputers~\cite{arute2019quantum,wu2021strong,morvan2023phase,decross2024computational}.  
However, such sampling tasks face inherent limitations: their outputs 
are difficult to verify, highly susceptible to noise, and challenging to reproduce 
across heterogeneous quantum hardware.  
As devices grow in scale and fidelity, there is a strong need for 
benchmarking tools that remain both experimentally scalable and 
verifiable, while still probing highly complex patterns 
of quantum dynamics.

Out-of-time-order correlators (OTOCs) 
have emerged as physically grounded and experimentally accessible probes 
of the complexity of Hamiltonian dynamics~\cite{larkin1969quasiclassical,shenker2014black,hosur2016chaos,maldacena2016bound,swingle2016measuring,roberts2017chaos}.  
Originally introduced in the context of quantum chaos and black-hole 
physics~\cite{larkin1969quasiclassical,shenker2014black,maldacena2016bound}, OTOCs quantify the sensitivity of quantum evolution to 
perturbations and capture how local information spreads throughout a 
quantum system~\cite{hosur2016chaos,roberts2017chaos}.  
Over the past decade, OTOCs have been measured in ultracold atoms~\cite{pegahan2021energy}, 
trapped ions~\cite{garttner2017measuring,landsman2019verified,joshi2020quantum}, NMR systems~\cite{li2017measuring,wei2018exploring}, and superconducting quantum processors~\cite{green2022experimental,braumuller2022probing}, 
revealing butterfly effects, operator spreading, and dynamical 
phase transitions.

Recent experimental advances have renewed and expanded the role of 
OTOCs in the characterization of quantum dynamics.  
Google Quantum AI reported a generalized family of higher-order OTOCs 
implemented using nested echo circuits on a large-scale superconducting 
processor, observing order-resolved constructive interference at the 
edge of ergodicity~\cite{google2025observation}.  
Concurrently, an NMR-based experiment demonstrated precision OTOC 
measurements using coherent multi-pulse control sequences~\cite{zhang2025quantum}, showcasing 
once again the exceptional controllability of NMR, a platform that played 
a foundational role in the earliest demonstrations of quantum information processing.

Indeed, many core ideas in quantum control~\cite{haeberlen2012high}, dynamical decoupling~\cite{viola1998dynamical,uhrig2011keeping}, and 
Hamiltonian engineering~\cite{waugh1968approach} were first developed in the context of NMR, and 
these techniques have had a profound and lasting influence on modern 
quantum algorithm design.  
A striking example of this intellectual lineage is the development of 
\emph{quantum signal processing} (QSP)~\cite{low2017optimal,low2019hamiltonian} and its matrix-valued extension, 
the \emph{quantum singular value transformation} (QSVT)~\cite{gilyen2019quantum,martyn2021grand}.  
These frameworks, rooted in the physics of coherent control and 
phase-modulated pulse sequences, provide optimal means of implementing 
polynomial transformations of eigenvalues or singular values using only 
single-qubit rotations and controlled reflections.  
QSP and QSVT now underpin state-of-the-art quantum algorithms for 
Hamiltonian simulation~\cite{low2019hamiltonian} and linear systems~\cite{gilyen2019quantum}.

Against this rich backdrop, a natural question arises: can the behavior of
OTOCs be given a precise operational understanding through the lens of
quantum algorithms?  
In this work, we show that the QSP framework provides exactly such an
algorithmic perspective, yielding a clear and constructive
operational understanding of higher-order OTOCs.
We prove that higher-order OTOCs implemented via echo sequences correspond 
exactly to Chebyshev polynomial transformations of the singular values of 
a \emph{spatially resolved truncated propagator} 
\(
A_{i,j}(t)=\bra{0_i}U(t)\ket{0_j}.
\)  
This perspective reveals that $\mathrm{OTOC}^{(k)}$ directly measures the 
$2k$-th Fourier component of the phase distribution associated 
with the singular values of $A_{i,j}(t)$. 
This is a clear answer to why time-ordered correlators (TOCs), i.e., the first harmonic often fail to 
detect causal-cone structures while higher-order OTOCs remain sharply 
sensitive to operator spreading and interference patterns.

We analyzed how the singular-value distributions and their associated phase distributions behave for several representative classes of dynamics including Haar-random unitaries, dynamics outside the causal cone, and free-fermion evolution.
These characteristic behaviors are further confirmed by numerical simulations, which demonstrate that higher-order OTOCs provide a sharp diagnostic of the arrival of the edge of ergodicity.

Building further on this connection, we introduce a generalized 
framework that we call \emph{OTOC spectroscopy}.  
By replacing Pauli flips in the echo sequence with tunable $Z$-rotations, 
we obtain arbitrary degree-$2d$ polynomial transformations of the 
singular values via depth-$d$ QSP sequences.  
This enables frequency-selective filters, such as low-pass, high-pass, or 
band-pass, acting directly on the dynamical spectrum of 
the singular values of truncated propagator
$A_{i,j}(t)$.  
Such filter-engineered OTOCs will allow us to isolate specific Fourier 
modes of the underlying phase distribution, providing a programmable and 
mode-resolved probe of scrambling, ergodicity, and spectral mixing in 
quantum many-body systems.
This OTOC spectroscopy not only explains long-standing features of 
OTOC behavior across chaotic, integrable, and localized regimes, but also 
opens new avenues for designing scalable, verifiable, and 
spectrally-resolved diagnostics of complexity of quantum many-body dynamics.

\section{QSP formulation of OTOC$^{(k)}$}
\label{sec:otoc-qsp}
\subsection{Higher order OTOC}
We consider an $N$-qubit system.  
Let $U$ be an $N$-qubit unitary, such as a time-evolution operator 
$U(t)=e^{-iHt}$, for which we want to know the behavior of dynamics.  
We pick two sites $i$ and $j$ and define single-qubit Hermitian unitary (e.g. Pauli) operators
$B_i$ and $M_j$ acting on the $i$th and $j$th qubits, respectively. 
The generalized OTOC echo operator is defined by~\cite{google2025observation}
\begin{equation}
    C_{i,j} := U^\dagger B_i U M_j.
\end{equation}
The $k$th-order OTOC between sites $i$ and $j$ is
\begin{equation}
    \mathrm{OTOC}^{(k)}_{i,j}(U,B_i,M_j)
    := \bra{\psi_{\rm ref}} C_{i,j}^{2k} \ket{\psi_{\rm ref}},
\end{equation}
for some reference state $\ket{\psi_{\rm ref}}$. In the following, we will focus on
$\ket{0^N}$ for concreteness.  
The exponent $2k$ arises naturally from the nested time-reversal structure
of the echo sequence.

When $k=1$, this expression reduces to the 
usual commutator-squared OTOC under standard assumptions; higher $k$ 
generalize the notion of operator spreading to multiple interference paths.
For $k=1/2$ and $B_i=M_i$, the quantity 
$\mathrm{OTOC}^{(1/2)}_{i,i}=\bra{\psi_{\rm ref}} C_{i,i}\ket{\psi_{\rm ref}}$ 
corresponds to a conventional time-ordered correlator (TOC).

\subsection{QSP formulation}
To connect this to a standard QSP setting with a single {\it probe} qubit, 
we introduce a SWAP gate $S_{j\leftrightarrow i}$ that exchanges site $j$
with a designated probe site $i$.  
We then define the modified unitary
\begin{equation}
    \tilde U := U S_{j\leftrightarrow i}.
\end{equation}
In the Heisenberg picture, this simply relabels the site on which $M$ acts, i.e.,
$M_i = S_{j\leftrightarrow i}M_jS_{j\leftrightarrow i} $.

We now regard the $i$th (probe) qubit as an ancilla used for 
block encoding.  
Write $\tilde U$ in block form with respect to the decomposition
$\mathcal{H}=\mathbb{C}^2\otimes\mathcal{H}_\text{sys}$:
\[
\tilde U =
\begin{pmatrix}
A & B' \\
C' & D
\end{pmatrix},
\qquad
A = \bra{0_i} \tilde U\ket{0_i},
\]
where $A$ acts on the $(N{-}1)$-qubit system register.
This is precisely a block-encoding of $A$.
From the viewpoint of the original unitary $U$, the operator $A$ is defined as 
\[
A_{i,j} =
\bra{0_i} U \ket{0_j}
\] 
such that the qubit indices of the truncation on the right and left sides differ. 
This distinction becomes important
because it allows one to incorporate its spatial structure, such as the distance between $i$ and $j$.
Therefore we call $A_{i,j}$ {\it spatially resolved truncated propagator}.

Let $\lambda_l$ be the singular values of $A$, and let $\ket{\psi_l}$ and 
$\ket{\phi_l}$ denote the corresponding left and right singular vectors:
\[
A\ket{\phi_l} = \lambda_l\ket{\psi_l}.
\]
By the qubitization~\cite{low2019hamiltonian,gilyen2019quantum}, the action of $\tilde U$ on the $2$-dimensional
invariant subspace associated with each $\lambda_l$ is given by a
$Y$-basis rotation:
\begin{equation}
    \tilde U_{\lambda_l} =  e^{-i (\theta_l/2) Y},
    \qquad
    \cos (\theta_l/2) = \lambda_l.
\end{equation}

For definiteness, take $B=M=Z$ acting on the probe qubit.  
Then
\[
C = \tilde U^\dagger Z_i \tilde U Z_i,
\]
and one finds that
\[
C_{\lambda_l} =  e^{i \theta_l Y},
\qquad
C_{\lambda_l}^{2k} = e^{i2k\theta_l Y}.
\]
Projection onto the probe state gives
\begin{align}
\bra{0_i} \bra{0^{N-1}} C^{2k} \ket{0_i} \ket{0^{N-1}}
&= \sum_l |\alpha_l|^2 \cos(2k\theta_l)
\\
&= \sum_l |\alpha_l|^2 T_{4k}(\lambda_l),
\end{align}
where $T_m$ is the Chebyshev polynomial of the first kind and
$\ket{\psi_{\rm ref}}=|0^N\rangle=\ket{0_i}\otimes\sum_l\alpha_l\ket{\psi_l}$.
Specifically, for the reference state $|0^N\rangle$, 
$\alpha_l = \langle \psi _l |0^{N-1} \rangle$.
Note that since we parametrize the singular values as $\lambda=\cos(\theta/2)$, the $2k$-th harmonic $\cos(2k\theta)$ corresponds to the $4k$-th order Chebyshev polynomial $\cos(2k\theta) = \cos(4k(\theta/2)) = T_{4k}(\lambda)$.

\begin{theorem}[OTOC as a Chebyshev polynomial transform]
Let $U$ be the given dynamics and take $B_i=Z_i$ and $M_j=Z_j$ on the probe qubit.
Then the $k$th-order OTOC associated with this echo sequence satisfies
\begin{align}
\mathrm{OTOC}^{(k)}(U,Z_i,Z_j)
&= \sum_l |\alpha_l|^2 \cos(2k\theta_l)
\\
&= \sum_l |\alpha_l|^2 T_{4k}(\lambda_l),
\end{align}
where $\lambda_l$ are the singular values of $A=\bra{0_i} \tilde U\ket{0_i}$ with $\tilde U = U S_{j\leftrightarrow i}$ (or equivalently $A_{i,j}= \bra{0_i} U\ket{0_j}$) and
$\ket{\psi_{\rm ref}}=\ket{0_i} \otimes\sum_l\alpha_l\ket{\psi_l}$.
\end{theorem}

\section{Behavior of OTOC$^{(k)}$ from the viewpoint of Fourier analysis}
\subsection{Phase distribution of singular values of spatially resolved truncated propagator}
From the above theorem, 
we see that the output of $\mathrm{OTOC}^{(k)}$ extracts a specific Fourier mode of the distribution function of the phase $\theta$ corresponding to the singular values as follows.
Let $\lambda_l^{(i,j)}$ be the singular values and define the
phase variables
\[
\lambda_l^{(i,j)}=\cos(\theta_l^{(i,j)}/2),
\qquad 
\theta_l^{(i,j)}(t)\in[0,\pi].
\]
Then the distribution function of $\theta$ can be given by
\begin{equation}
  p_{i,j}(\theta)
  := \sum_{l}
     |\alpha_l ^{(i,j)}|^2\,
     \delta\!\left(\theta-\theta_l ^{(i,j)}\right).
\end{equation}
Now we realize 
\[
\mathrm{OTOC}^{(k)}(U,Z_i,Z_j) = \int _0 ^{\pi} d\theta  \cos (2k \theta) p_{i,j}(\theta) .
\]
$\mathrm{OTOC}^{(k)}$ corresponds to the $2k$-th Fourier mode with $\cos (2k \theta)$.
Specifically, TOC with $k=1/2$ corresponds to the first harmonic with $\cos (\theta)$.

Although the phase distribution $p_{i,j}(\theta)$ depends 
on the chosen reference state, the underlying distribution of 
$\theta$ associated with the truncated unitary operator 
$A_{i,j} $ is determined solely by the dynamics $U$ and two sites $i$ and $j$.
In the following, we analyze how the behavior of this intrinsic 
phase distribution $\tilde p_{i,j}(\theta)$ (or singular value distribution) of $A_{i,j}$ governs the detectability of different 
Fourier modes (Chebyshev moments) and clarifies the contrasting responses of TOCs and 
higher-order OTOCs.

\subsection{Haar random unitary}

When $U$ is Haar random, $\tilde U = U S_{j\leftrightarrow i}$ is also Haar random.
Then the spatially resolved truncated propagator 
\[
A_{i,j}=\bra{0_i}\tilde U\ket{0_i}
\]
is a \emph{truncated Haar unitary}~\cite{zyczkowski2000truncations,collins2005product,dong2012circular}.
For such truncations, it is well known that the eigenvalues 
$x=\lambda^2$ of $A_{i,j}A_{i,j}^\dagger$ follow the Jacobi (or arcsine) ensemble
 with $\alpha=\beta =-1/2$ in the limit of large $N$:
\begin{equation}
\rho_x(x) = \frac{1}{\pi\sqrt{x(1-x)}}, \qquad x\in(0,1),
\end{equation}
where $\lambda$ denotes the singular values of $A_{i,j}$.

Changing variables from $x=\lambda^2$ to the singular value
$\lambda\in(0,1)$ yields
\begin{equation}
\rho_\lambda(\lambda)
= \rho(\lambda^2)\,2\lambda
= \frac{2}{\pi\sqrt{1-\lambda^2}}, 
\qquad 0<\lambda<1.
\end{equation}
Finally, using the phase parametrization
\[
\lambda=\cos\frac{\theta}{2},\qquad \theta\in(0,\pi),
\]
the corresponding distribution of $\theta$ becomes
\begin{equation}
\rho_\theta(\theta)
= \rho_\lambda(\cos(\theta/2))
\left|\frac{d\lambda}{d\theta}\right|
= \frac{1}{\pi},
\qquad 0<\theta<\pi.
\end{equation}
Thus the intrinsic phase distribution of the truncated propagator is 
perfectly uniform on $[0,\pi]$ in the Haar-random limit.

As a consequence, all nontrivial Fourier components vanish:
\[
\int_0^\pi \frac{1}{\pi} \cos(n\theta)\, d\theta = 0,
\qquad n\ge 1,
\]
and therefore both TOCs and higher-order OTOCs, 
$\mathrm{OTOC}^{(k)}$ decay to zero.
This Haar-random distribution provides a natural reference against which 
nontrivial dynamical features, such as causal-cone edges, interference 
patterns, or localization, appear as deviations from the flat phase 
distribution and its Fourier harmonics.

\subsection{Outside the causal-cone}
To gain intuition for the opposite extreme, 
let us consider the case $U=I$.
The truncated propagator reduces to
\[
A_{i,j}= \ket{0_j} \otimes \bra{0_i} \otimes I _{m \neq i,j} ,
\]
where $I _{m \neq i,j} $ indicates the identity operator on the qubits except for $i$th and $j$th.
As a result, the singular values of $A_{i,j}$ take the exact bimodal form
\[
\lambda \in \{0,\,1\},
\qquad
\text{with multiplicities }2^{N-2}\text{ each},
\]
corresponding respectively to non-propagating directions and 
projection-induced local directions.

Next, let us consider the case where $i$th site lies outside the causal cone of $j$th site.
More precisely, we consider
a short time scale $t < d(i,j)/v_{\rm {LR}}$
with $d(i,j)$ being the distance between $i$ and $j$,
where the Lieb–Robinson bound ensures a correlation between
$i$ and $j$ is not yet formed.
The singular values of $A_{i,j}(t)$ are the square roots of the eigenvalues of
\begin{equation}
    C_{i,j}(t) := A_{i,j}(t) A_{i,j}(t)^\dagger
    = \langle 0_i | U(t) P_j U^\dagger(t) | 0_i \rangle,
\end{equation}
where $P_j := |0_j\rangle\!\langle 0_j|$.
Thus the problem reduces to understanding the spectrum of the positive operator
$C_{i,j}(t)$.

When $i$th and $j$th sites are outside of the causal cone,
the Heisenberg-evolved projector
\begin{equation}
    P_j(t) := U(t) P_j U^\dagger(t)
\end{equation}
has support only on sites within a finite neighborhood of $j$, and in
particular acts as the identity on site $i$.
Equivalently, $P_j(t)$ commutes with the local projector
$P_i := |0_i\rangle\!\langle 0_i|$,
\begin{equation}
    [P_i, P_j(t)] \simeq 0 \qquad
    \text{for } t < d(i,j)/v_{\mathrm{LR}},
\end{equation}
with $v_{\mathrm{LR}}$ being the Lieb–Robinson velocity.
Hence $C_{i,j}(t)$ is almost unitarily equivalent to $C_{i,j}(0)$, and its spectrum
is almost identical to the spectrum in the trivial case $U(t)=I$.

The above discussion implies that even when the time $t$ is small and $j$ lies outside the causal-cone of $i$, the left and right truncations commute, resulting in the same bimodal distribution. 
Only when the causal-cone reaches the separation between $i$ and $j$ do 
the left and right truncations begin to communicate with overlap, after which the two-peak 
structure rapidly dissolves.  
In fully chaotic dynamics, the ensuing spectral mixing drives 
$A_{i,j}(t)$ towards a truncated-Haar-like regime in which the phase 
distribution becomes nearly uniform, as will be corroborated by our 
numerical simulations.

\subsection{Algorithmic meanings of higher order OTOC}
This dichotomy between a bimodal distribution (outside the causal cone) 
and a nearly uniform distribution (inside the cone under chaotic 
dynamics) has direct consequences for different Fourier modes.  
The TOC-like quantity corresponds to the first harmonic 
$\cos\theta$, but for a bimodal distribution supported at 
$\theta=0$ and $\theta=\pi$, we have 
\[
\cos 0 = +1,
\qquad
\cos\pi = -1,
\]
so that the contributions cancel almost exactly.  
Thus, the first Fourier component of a bimodal distribution is nearly 
indistinguishable from that of the uniform (Haar) distribution,  
explaining why TOCs fail to detect the presence of a causal-cone edge.

In contrast, the second harmonic $\cos(2\theta)$, probed by 
$\mathrm{OTOC}^{(1)}$ in our notation, takes the value $+1$ at both 
$\theta=0$ and $\theta=\pi$, whereas it averages to zero under a uniform 
distribution.  
Hence OTOC$^{(1)}$ is sharply sensitive to the appearance of 
a bimodal spectrum and provides a clear discriminator between  
non-propagating, projection-dominated behavior outside the causal 
cone and  
ergodic spectral mixing inside the cone.

Higher harmonics $\cos(4\theta),\cos(6\theta),\dots$, corresponding to 
OTOC$^{(k)}$ with $k\ge 2$, amplify this contrast even further, as their 
oscillatory structure selectively accentuates sharp peaks near the 
endpoints $\theta=0,\pi$.  
Thus higher-order OTOCs act as progressively finer ``spectral filters'' 
that capture the detailed structure of the truncated propagator's 
phase distribution.

\subsection{Structural signature of underlying Hamiltonian}
Next, let us examine how the phase distribution changes depending on the
properties of the Hamiltonian.
Suppose the dynamics $U$ is given as a Hamiltonian evolution $U=e^{-iHt}$
with respect to a Hamiltonian $H$.
The truncated unitary is now defined depending on a time $t$:
\[
A_{i,j} (t) = \bra{0_i}  e^{ -i Ht } \ket{0_j}.
\]
The singular values of the truncated propagator 
$A_{i,j}(t)$ quantify how strongly the time evolution 
transfers components of the input subspace conditioned on $\ket{0_j}$ to the
output subspace conditioned on $\bra{0_i}$.  
More precisely, if 
$A_{i,j}(t)\ket{\phi_l^{(j)}}=\lambda_l(t)\ket{\psi_l^{(i)}}$ 
is the singular-value decomposition, then each 
$\lambda_l(t)\in[0,1]$ measures the amplification (or more precisely attenuation since $\lambda \leq 1$) 
with which the input direction $\ket{\phi_l^{(j)}}$ is mapped into 
the output direction $\ket{\psi_l^{(i)}}$.  
When $U$ is close to the identity or when $i$ and $j$ lie outside each 
other's causal cone, most input directions experience almost complete 
suppression, yielding $\lambda_l(t)\simeq 0$, while a few 
projection-induced local directions may satisfy 
$\lambda_l(t)\simeq 1$.  
Conversely, when coherent flow from $j$ to $i$ builds up, the largest 
singular values approach unity.  
Thus the singular-value distribution of $A_{i,j}(t)$ directly encodes 
the structure and strength of information transfer between the two 
local subspaces.

A complementary viewpoint is obtained by inserting the spectral 
decomposition 
$U(t)=\int e^{-i\omega t}E_H(d\omega)$, where the operator-valued distribution function 
$E_H(d\omega)$ is defined as follows:
\[
E_H(d\omega) 
:= \sum_{a} \delta(\omega - E_a)\, |E_a\rangle\langle E_a|\, d\omega.
\]  
This yields 
\[
A_{i,j}(t)
= \int e^{-i\omega t}\, \bra{0_i} E_H(d\omega)\ket{0_j} ,
\]
revealing that the truncated propagator is a coherent superposition of 
energy-resolved transfer channels between $j$ and $i$.  
Different energy sectors $\omega$ acquire phase factors 
$e^{-i\omega t}$, and their constructive or destructive interference, 
arising from energy differences $\omega-\omega'$, determines the growth, 
suppression, and temporal modulation of the singular values.  
Indeed, this can be understood explicitly by considering
\begin{align}
&A_{i,j}(t)A_{i,j}^\dagger(t)
 \\
 &=\iint 
   e^{-it(\omega-\omega')}
   \, \bra{0_i} E_H(d\omega) \ket{0_j} \bra{0_j} E_H(d\omega') \ket{0_i}.
\end{align}
The eigenvalues of $A_{i,j}(t)A_{i,j}^\dagger(t)$,
whose square roots correspond to the singular values of $A_{i,j}(t)$, 
are shaped by the interference 
kernel $e^{-it(\omega-\omega')}$ coupling different energy sectors.
When $\omega\neq\omega'$, rapid oscillations of this phase factor 
suppress off-diagonal contributions and drive the spectrum toward a 
mean-field, Haar-like form. On the other hand, when $\omega$ and $\omega'$ are aligned 
(e.g., due to integrability or localization), coherent reinforcement 
stabilizes sharp structures in the singular-value distribution.

\subsection{Free Fermion example}
As an exercise, let us consider a free-fermionic Hamiltonian given by 
\begin{equation}
    H = \sum_{a,b} h_{ab} \hat{c}_a^\dagger  \hat{c}_b ,
\end{equation}
where $\hat{c}_a$ are fermionic annihilation operators and $h$ is an
$N\times N$ single-particle hopping matrix.
The many-body time evolution $U(t)=e^{-iHt}$ is a Gaussian unitary, and the
Heisenberg evolution of $\hat{c}_a$ is governed by the single-particle propagator
\begin{equation}
    U^\dagger(t) \hat{c}_a U(t)
    = \sum_b G_{ab}(t)\, \hat{c}_b,
    \qquad
    G(t) = e^{-i h t}.
\end{equation}
The local projector $P_j = |0_j\rangle\!\langle 0_j| = 1 - \hat{c}_j^\dagger \hat{c}_j$
evolves as
\begin{equation}
    U(t) P_j U^\dagger(t)
    = 1 - \hat{c}_j^\dagger(t)\hat{c}_j(t),
    \qquad
    \hat{c}_j(t) = \sum_a G_{ja}(t) \hat{c}_a.
\end{equation}

We now project site $i$ onto $|0_i\rangle$.
Writing
\begin{equation}
    \hat{c}_j(t)
    = G_{ji}(t) \hat{c}_i + \sum_{a\neq i} G_{ja}(t) \hat{c}_a,
\end{equation}
and defining the vector $v$ on the remaining modes by
\begin{align}
    v_a(t) &:= G_{ja}(t)\quad (a\neq i), 
    \\
    \|v(t)\|^2 &= \sum_{a\neq i} |v_a(t)|^2
                = 1 - |G_{ji}(t)|^2 ,
\end{align}
we introduce a normalized fermionic mode
\begin{equation}
    \hat{c}_v := \frac{1}{\|v\|} \sum_{a\neq i} v_a\, \hat{c}_a ,
\end{equation}
so that $\hat{c}_j(t) = G_{ji}(t) \hat{c}_i + \|v(t)\|\, \hat{c}_v$.
Substituting into $U(t)P_jU^\dagger(t)$ and taking the expectation value in
$|0_i\rangle$, all terms containing $\hat{c}_i$ or $\hat{c}_i^\dagger$ drop out, yielding
\begin{align}
    C_{i,j}(t)
    &= \langle 0_i| U(t) P_j U^\dagger(t) |0_i\rangle
    = 1 - \|v(t)\|^2\, \hat{c}_v^\dagger \hat{c}_v
    \\
    &= 1 - \bigl(1-|G_{ji}(t)|^2\bigr)\, \hat{n}_v,
\end{align}
with $\hat{n}_v := \hat{c}_v^\dagger \hat{c}_v$ the number operator for the single mode $\hat{c}_v$.
Since $\hat{n}_v$ has eigenvalues $0$ and $1$, the spectrum of $C_{i,j}(t)$ is
\begin{equation}
    \lambda_0 = 1, \qquad
    \lambda_1 = |G_{ji}(t)|^2,
\end{equation}
each with degeneracy $2^{N-2}$ corresponding to the occupations of the other
$N-2$ fermionic modes.
Therefore, the singular values of the truncated propagator $A_{i,j}(t)$ are $1$ and $|G_{ji}(t)|$,
where $|G_{ji}(t)|^2$ is the single-particle
propagation probability from $j$ to $i$.

\subsection{Implications of Fourier mode analysis}
On the other hand, in chaotic systems, the broad and irregular energy spectrum leads to 
rapid dephasing across sectors, driving the phase distribution 
$p_{i,j}(\theta,t)$ toward a nearly uniform form characteristic of 
truncated Haar random matrices.  
In integrable systems, the quasiparticle structure of the spectrum 
induces persistent oscillatory patterns in $\tilde p_{i,j}(\theta,t)$, while in 
many-body localized systems the spectral weight remains concentrated 
near $\theta\simeq 0$, reflecting the stability of local integrals of 
motion.  
OTOCs provide a natural tool to probe these differences: 
$\mathrm{TOC}$ corresponds to the first Fourier mode, which nearly 
vanishes for both the bimodal distributions encountered outside the 
causal cone and the uniform distributions generated by chaotic mixing.  
In contrast, $\mathrm{OTOC}^{(1)}$ and higher-order OTOC$^{(k)}$ extract 
higher Fourier harmonics $\cos(2k\theta)$ of $\tilde p_{i,j}(\theta,t)$, making 
them sharply sensitive to the presence of coherent peaks at 
$\theta=0,\pi$ as well as to finer interference structures arising from 
the underlying dynamics.  
This Fourier-mode viewpoint clarifies why higher-order OTOCs serve as 
powerful diagnostics of operator spreading and spectral mixing depending on the structure of given Hamiltonian.
Chaotic, integrable, and many-body localized (MBL) dynamics produce
distinct behaviors: chaotic dynamics lead to broad, nearly uniform
distributions; integrable dynamics produce sharp, oscillatory features; 
and MBL dynamics yield distributions localized near $\lambda\approx 1$
for $i=j$ and rapidly decaying with $|i-j|$.
These features will be verified with numerical simulations later.

\section{OTOC spectroscopy via QSP}
\label{sec:qsp-otoc}
Building on the understanding that the QSP formulation of 
$\mathrm{OTOC}^{(k)}$ measures the $2k$-th Fourier mode of the singular-value 
phase distribution, it is natural to generalize the construction by 
replacing the fixed $Z$ flip in QSP with $Z$-rotations $e^{-i \phi Z}$.  
This introduces a transformation of the singular value of $A_{i,j}$, allowing us to 
form linear combinations of Fourier modes and thereby implement 
frequency-selective filter functions.  
As a natural extension of $\mathrm{OTOC}^{(k)}$, we will refer to this 
framework as \emph{OTOC spectroscopy} in the following.

To make this generalization explicit, we define single-qubit phase
rotations on sites $i$ and $j$ as
\[
B_i(\phi) := e^{-i\phi Z_i},
\qquad 
M_j(\phi) := e^{-i\phi Z_j}.
\]
For each layer $r=1,\dots,d$, we introduce the generalized echo operator
\begin{equation}
C^{(r)}_{i,j}(t)
    := U^\dagger(t)\, B_i(\phi_{2r+1})\, 
       U(t)\, M_j(\phi_{2r}),
\label{eq:generalized_echo}
\end{equation}
where the phases $\vec{\phi} = \{\phi_r\}$ are free parameters to be chosen according
to the desired polynomial transformation.

Using these building blocks, we define the $d$-layer \emph{QSP-OTOC}
corresponding to a phase sequence $\vec{\phi}=(\phi_0,\phi_1,\dots,\phi_{2d})$ as
\begin{equation}
\mathrm{QSP\mbox{-}OTOC}^{(d)}_{i,j}(t,\vec{\phi})
:= 
\bra{0^N}\,
B_i(\phi_{2d})\,
\prod_{r=1}^{d} 
C^{(r)}_{i,j}(t)
\,\ket{0^N}.
\label{eq:qsp-otoc-definition}
\end{equation}
For special choices of $\vec{\phi}$,
this expression reduces to the standard higher-order OTOCs discussed 
above, where the action on each singular value 
$\lambda=\cos(\theta/2)$ is given by the Chebyshev polynomial 
$T_{4k}(\lambda)$ with $k=d$.  
In this sense, $\mathrm{QSP\mbox{-}OTOC}^{(d)}$ provides a natural 
generalization of $\mathrm{OTOC}^{(k)}$ in which the polynomial order 
and structure are controlled by the freely tunable phase sequence.

More generally, the sequence~\eqref{eq:qsp-otoc-definition} 
implements, on each singular-value sector $\lambda_l^{(i,j)}(t)$ of 
$A_{i,j}(t)$, a polynomial transformation
\[
\lambda_l \;\mapsto\; P_{d,\vec{\phi}}(\lambda_l),
\]
where $P_{d,\vec{\phi}}(\lambda)$ is a degree-$2d$ polynomial obtained 
through the standard QSP synthesis applied to the walk operator 
associated with $A_{i,j}(t)$.  
Thus,
\begin{equation}
\mathrm{QSP\mbox{-}OTOC}_{i,j}^{(d)}(t,\vec{\phi})
= 
\sum_l |\alpha_l^{(i,j)}|^2\,
P_{d,\vec{\phi}}\!\left(\lambda_l^{(i,j)}(t)\right).
\label{eq:qsp-otoc-polynomial}
\end{equation}
Since 
any real polynomial of degree at most $2d$ satisfying the standard QSP parity and boundedness constraints
can be realized by an 
appropriate choice of phases $\vec{\phi}$, the QSP-OTOC implements 
a linear combination of Fourier harmonics  of the phase distribution
$\tilde p_{i,j}(\theta,t)$ that lies within the class of 
bounded degree-$2d$ polynomials implementable by QSP.
This enables the design of frequency-selective \emph{filters} acting on 
the phase spectrum of $A_{i,j}(t)$.
For instance, low-pass, high-pass, and band-pass filters can be designed 
to detect a characteristic feature of the Hamiltonian dynamics of interest.
In chaotic or Haar-random regimes, where $p_{i,j}(\theta,t)$ approaches 
a nearly uniform distribution, such filters enable a refined diagnosis 
of residual structure and deviations from randomness.  
Conversely, in integrable or many-body localized dynamics, where 
$p_{i,j}(\theta,t)$ exhibits sharp spectral support, QSP-OTOCs 
provide a powerful means to resolve the underlying interference 
structure by selectively probing specific harmonics of the phase 
distribution.

In summary, the freedom to engineer $P_{d,\vec{\phi}}(\lambda)$ through 
QSP transforms the OTOC protocol into a fully programmable spectroscopic 
tool, which allows us to isolate dynamical features that are invisible to 
fixed-order OTOCs and greatly extending the versatility of 
out-of-time-order diagnostics.

\section{Numerical Experiments}
\label{sec:numerics}

To substantiate the QSP-based interpretation developed above, we perform
numerical simulations to see the singular value distributions of the truncated propagators $A_{i,j}(t)$ 
and time-dependent behavior of their Fourier modes (for the corresponding phase distribution)
under several physically relevant many-body Hamiltonians.  

\subsection{Models}
We consider three benchmark one-dimensional spin chains:
\\
{\it Chaotic (ergodic) model:}
      a XYZ chain with a magnetic field~\cite{shiraishi2019proof}
      \[
      H_{\mathrm{chaos}}
      = \sum_i (J_x X_iX_{i+1}+ J_y Y_iY_{i+1}+ J_zZ_iZ_{i+1})
      + h \sum_i  Z_i ,
      \]
      where we specifically choose $J_x = -0.4$, $J_y=-2.0$, $J_z=-1.0$ and 
      $h = 0.75$.
\\
{\it Integrable model:}
      an XXZ chain with uniform field
      \[
      H_{\mathrm{int}}
      = \sum_i J (X_iX_{i+1}+Y_iY_{i+1}+\Delta Z_iZ_{i+1})
      + h\sum_i Z_i,
      \]
where we take $J=1$, $\Delta=0$, and $h=0$ as a free-fermion case.
We also consider $J=1$, anisotropy $\Delta=1$, and a uniform field $h=0.2$, which yields an interacting integrable model solvable by the Bethe ansatz.
\\

{\it MBL model:}
      a strongly disordered Heisenberg chain
      \[
      H_{\mathrm{MBL}}
      = \sum_i J (X_iX_{i+1}+Y_iY_{i+1}+Z_iZ_{i+1})
      + \sum_i h_i Z_i ,
      \]
      where $J=1$ and the on-site fields $h_i$ 
      are drawn uniformly at random from
      the interval $[-h,h]$ with $h=5$, ensuring
      $|h_i|\gg J$ and hence localization.

System sizes $N=10$ are diagonalized exactly.  
Time evolution is computed via
\(
U(t)=e^{-iHt}
\)
using standard exponentiation routines.
For any pair of sites $i,j$, the truncated propagator is extracted as
\[
A_{i,j}(t) = \bra{0_i} U(t) \ket{0_j}.
\]
This operator is of dimension $2^{N-1}\times2^{N-1}$ and its singular
values $\lambda_l^{(i,j)}(t)$ are obtained via singular value 
decomposition.

We study the time evolution of the intrinsic phase distributions of the singular values:
\begin{align}
\tilde p_{i,j}(\theta,t)
    &= \frac{1}{D}\sum_l
       \delta\!\bigl(\theta-\theta_l^{(i,j)}(t)\bigr),
\end{align}
with $D$ being the dimension,
where $\theta_l^{(i,j)}(t) = 2\arccos (\lambda_l^{(i,j)}(t))$.
Furthermore,
we also observe $k$th order Chebyshev mode of 
the distribution of the singular values $\lambda_l^{(i,j)}(t)$:
\[
\mathcal{M}_{4k}^{(i,j)}(t)
    = \frac{1}{D}\sum_l
       T_{4k}\!\left(\lambda_l^{(i,j)}(t)\right),
\]
for $4k=2,4,8, 12$, each of which corresponds to TOC, OTOC$^{(1)}$, OTOC$^{(2)}$ and OTOC$^{(3)}$,
respectively.

\subsection{Results}
\begin{figure*}
    \centering
    \includegraphics[width=1.0\linewidth]{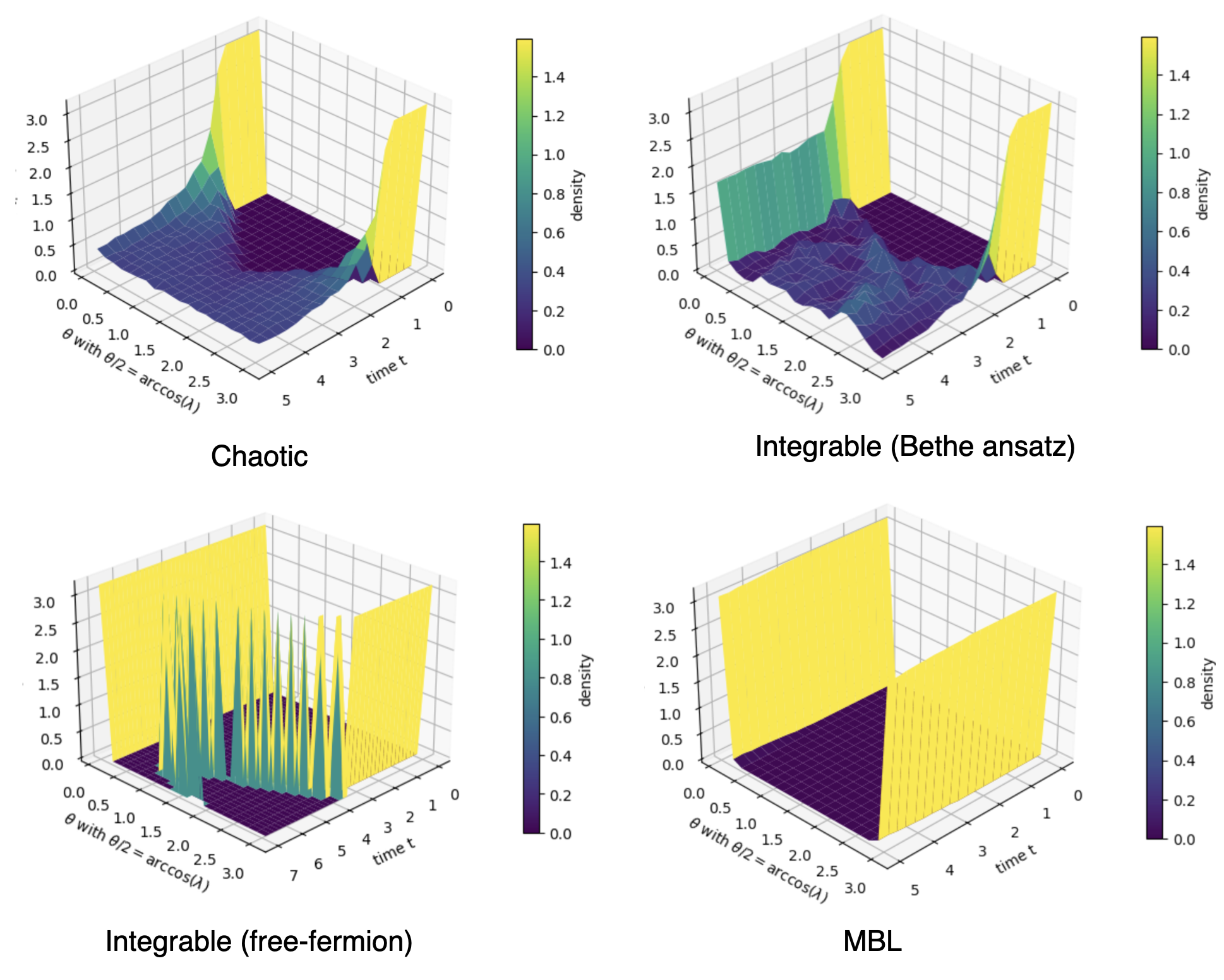}
    \caption{
Phase distributions $\tilde p_{0,9}(\theta,t)$ of the singular values of
the truncated propagator $A_{0,9}(t)$ for chaotic (left top), integrable
(Bethe ansatz) (right top), integrable (free-fermion) (left bottom) and MBL (right bottom) dynamics.  The horizontal axes show time $t$  ($0\leq t \leq 7$ for integrable (free-fermion) and $0\leq t \leq 5$ for others) 
and the phase variable $\theta$ with $\lambda=\cos(\theta/2)$, and the
vertical axis and color scale indicate the density.
}
    \label{fig01}
\end{figure*}
Figure~\ref{fig01} illustrates the full time evolution of the
phase distribution $\tilde p_{i,j}(\theta,t)$ associated with the singular
values of the spatially resolved truncated propagator $A_{i,j}(t)$ for
chaotic, integrable, and MBL dynamics,
with $i=0$ and $j=9$.
The four panels correspond, from top left to bottom right, to the chaotic,
integrable (Bethe ansatz), integrable (free-fermion), and MBL chains,
respectively, providing a direct visualization of the causal structure and the
qualitative nature of the underlying many-body dynamics.

At the initial time $t=0$, when $U(t)=I$, the truncated propagator
$A_{i,j}(0)=\bra{0_i}I\ket{0_j}$ exhibits an exactly bimodal singular-value
spectrum: all weight is concentrated at $\theta=0$ and $\theta=\pi$,
corresponding respectively to the local projection channel and the
completely suppressed nonlocal channel.  
This yields a phase distribution $\tilde p_{i,j}(\theta,0)$ with two sharp peaks
at the endpoints of the interval $[0,\pi]$.

As time evolves, the behavior of the phase distribution changes
dramatically once the influence of site $j$ reaches site $i$.  
In both the chaotic and integrable (Bethe ansatz) chains, 
once site $i=0$ enters the causal cone of site $j=9$ between $t=1$ and $t=2$,
the initial bimodal structure collapses.
Beyond this point, the subsequent evolution reflects the qualitative
nature of the underlying dynamics.  
In the chaotic chain, the distribution quickly relaxes toward a broad,
nearly featureless profile approaching the truncated-Haar limit, with 
no residual oscillation in $\theta$.  
In contrast, the integrable (Bethe ansatz) chain displays pronounced and persistent
oscillatory patterns: spectral weight flows coherently across $\theta$,
producing the characteristic interference fringes associated with stable
quasiparticle modes.
This is more evident for the integrable (free-fermion) chain, where the initial peak at $\theta = \pi$ begins to oscillate once the causal cone reaches the site $j = 9$.
This behavior is completely consistent with the analytical argument given for the free-fermion case above.

The behavior in the MBL chain is strikingly different.  
Because localization prevents ballistic propagation, site $i$ remains
effectively outside the causal cone of site $j$ for all accessible
times.  
As a result, the phase distribution retains its initial bimodal form,
with weight pinned near $\theta\simeq 0$ and $\theta\simeq \pi$ and no
evidence of the spectral mixing seen in the chaotic and integrable
cases.

These observations demonstrate that the complexity of the dynamics ---
ergodic, integrable (free-fermion, Bethe ansatz), or localized --- is vividly encoded in the
singular-value statistics of the spatially resolved truncated
propagator.  
The phase distribution $\tilde p_{i,j}(\theta,t)$ provides a direct,
mode-resolved signature of information spreading and clearly delineates
the qualitative differences between chaotic mixing, coherent
quasiparticle propagation, and localization-protected dynamics.

\begin{figure*}
    \centering
    \includegraphics[width=\linewidth]{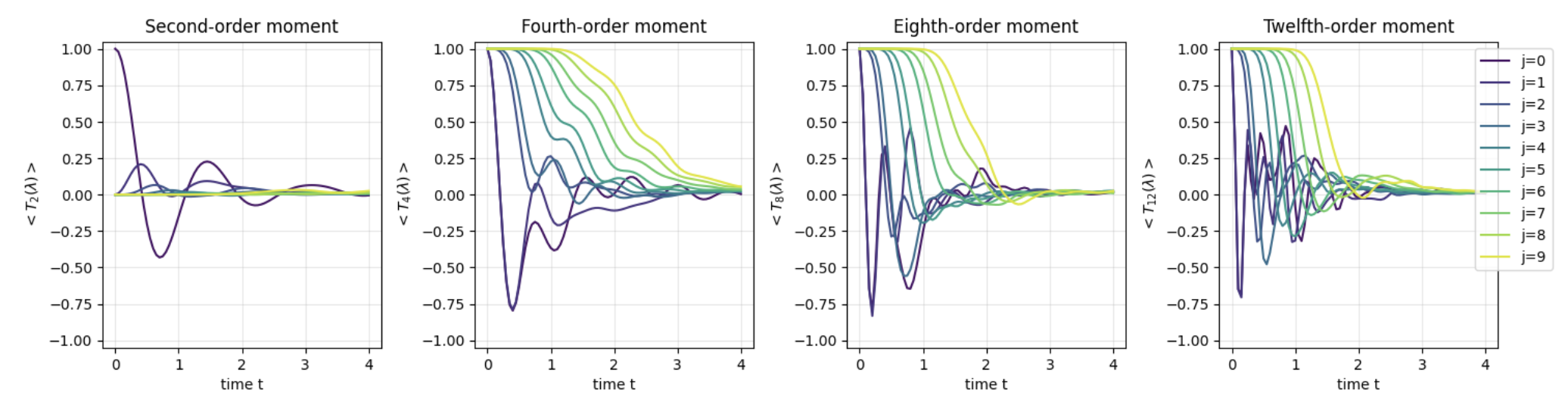}
    \caption{
Time evolution of the second-, fourth-, eighth-order, and twelfth-order Chebyshev
moments $\langle T_{4k}(\lambda)\rangle$ of the truncated propagator
$A_{i,j}(t)$ for the chaotic Hamiltonian, corresponding respectively to
the TOC ($k=1/2$), OTOC$^{(1)}$ ($k=1$), OTOC$^{(2)}$ ($k=2$), and OTOC$^{(3)}$ ($k=3$)
from left to right.
Each curve shows the moment for a different separation $j=0,...,9$ where $i=0$.
}
    \label{fig02}
\end{figure*}

Figure~\ref{fig02} displays the time evolution of the
Chebyshev moments $\langle T_{4k}(\lambda)\rangle$ of the singular-value
distribution of the truncated propagator $A_{i,j}(t)$ under the chaotic
Hamiltonian, for $k=1/2$ (TOC), $k=1$ (OTOC$^{(1)}$), $k=2$
(OTOC$^{(2)}$), and $k=3$ (OTOC$^{(3)}$).  
Each curve corresponds to a different separation~$j$, allowing us to
track how the spatially resolved moments change as site $i$ enters the
causal cone of site $j$.

In the leftmost panel ($k=1/2$), the curve for $j=0$ reproduces the
standard TOC, which decays monotonically from its initial value of~$1$.
For $j\neq 0$, however, the spatially resolved TOC does not clearly
detect the arrival of the causal-cone edge.
As explained earlier, the second-order Chebyshev moment $T_2(\lambda)$
corresponds to the first Fourier harmonic $\cos\theta$, which takes
opposite signs at $\theta=0$ and $\theta=\pi$.
Consequently, a bimodal distribution and a nearly uniform distribution
yield similar averaged signals, making the TOC largely insensitive to
whether $(i,j)$ lies inside or outside the causal cone.
Indeed, for all $j\neq 0$, the TOC signal rises from zero and settles to a
small value (approaching $0$ in the Haar-random limit), without a sharp
signature of the causal-cone crossing.

In the second panel ($k=1$), the moment corresponds to the second Fourier
harmonic $\cos(2\theta)$, i.e., the fourth-order Chebyshev polynomial $T_4$.
Although $\cos(2\theta)$ takes the value $+1$ at $\theta=0$ and
$\theta=\pi$, it has relatively broad support there.
As a result, when the initially bimodal phase distribution begins to deform
at early times—before becoming completely flat—the averaged value of
$\cos(2\theta)$ decays only gradually.
Thus the resulting OTOC$^{(1)}$ does not sharply detect the
causal-cone arrival. 

A sharp distinction first emerges for $k=2$ (third panel), corresponding to
$\cos(4\theta)$ and the eighth-order Chebyshev polynomial $T_8$.
The higher harmonic strongly amplifies narrow peaks near $\theta=0$ and
$\theta=\pi$, making the averaged signal highly sensitive to even small
departures from a bimodal distribution.
Consequently, the moment exhibits a sharp and rapid decay when the
causal cone reaches site $j$, cleanly resolving the light-cone structure
and giving a clear spectral signature of operator spreading.

Finally, for $k=3$ (rightmost panel), corresponding to $\cos(6\theta)$ and
the twelfth-order Chebyshev polynomial $T_{12}$, the behavior becomes even sharper.
The sixth harmonic further narrows the effective support around
$\theta=0$ and $\theta=\pi$, dramatically enhancing sensitivity to fine
structure in the phase distribution.
Thus OTOC$^{(3)}$ shows an even more pronounced, nearly discontinuous drop
at the causal-cone arrival time.

Taken together, these numerical observations are fully consistent with
the theoretical picture developed in this work:  
TOCs probe only the first Fourier mode and thus fail to resolve the
transition from bimodal to mixed phase distributions, whereas
higher-order OTOC$^{(k)}$ extract higher harmonics that sharply reveal the
onset of scrambling and the structure of the causal cone.  
These results substantiate our claim that OTOC$^{(k)}$ and QSP-based
OTOC spectroscopy provide mode-resolved diagnostics of many-body
dynamics that go well beyond conventional OTOC measurements.

\section{Conclusion and Discussion}

We have demonstrated that OTOCs admit a natural and powerful reformulation 
within the framework of quantum signal processing.  
Higher-order OTOCs correspond to Chebyshev polynomial transformations of the 
singular values of spatially resolved truncated propagators $A_{i,j}(t)$.  
In the fully chaotic (Haar random) limit, the singular values follow an arcsine law 
and all nontrivial Fourier components of the eigenphase distribution vanish, 
explaining the decay of $\mathrm{OTOC}^{(k)}$.

For Hamiltonian dynamics, however, the singular-value distributions 
or corresponding phase distributions retain rich structure reflecting locality, causal-cone 
propagation, and integrability or localization.  
We have shown that TOCs probe only first moments of self blocks $A_{i,i}(t)$ 
and therefore lose dynamical information rapidly, whereas OTOC and more 
general QSP-OTOCs probe higher Chebyshev moments of two-point blocks 
$A_{i,j}(t)$ and thus remain sensitive to operator spreading and coherent 
interference patterns deep into the scrambling regime.

By generalizing Pauli operations to tunable rotations, QSP-OTOCs implement 
arbitrary polynomial filters on the singular spectrum of $A_{i,j}(t)$.  
This enables new forms of ``OTOC spectroscopy,'' in which slow modes, 
fast modes, and intermediate interference structures can be selectively 
amplified by appropriate choices of $P(\lambda)$.  
We expect this perspective to be useful not only for interpreting current 
quantum echo experiments, but also for designing new protocols that probe 
many-body dynamics in a mode-resolved fashion with full spatial resolution.

\begin{acknowledgments}
KF would like to thank Kaoru Mizuta and Tatsuhiko N. Ikeda for valuable and insightful discussions. 
This work is supported by MEXT Q-LEAP Grant
No. JPMXS0120319794, JST CREST JPMJCR24I3, and JST COI-NEXT No.
JPMJPF2014.
\end{acknowledgments}
\appendix

\bibliographystyle{apsrev4-1}
\bibliography{references}

\end{document}